  \newcommand{\referee}[1]{#1}
 \newcommand{\camera}[1]{\relax}

\referee{
\documentclass[12pt]{article}
\usepackage{graphicx}
\newcommand{\etal}{{\it et al.}}
\newcommand{\tabla}{table 1}
\setlength{\textwidth}{15.5cm}
\setlength{\oddsidemargin}{0.5cm}
\setlength{\textheight}{21.5cm}
\setlength{\topmargin}{-1cm}
\setlength{\parskip}{12pt}
}

\usepackage{marvosym}
\usepackage{graphicx}

\begin{document}

\setlength{\unitlength}{0.9em}
\newcommand{\circulo}{
\begin{picture}(0.4,0.5)(-0.2,-0.3)
\circle{0.75}
\end{picture}
}
\newcommand{\cuadrado}{
\frame{
\begin{picture}(0.65,0.65)
\end{picture}}
}

\newcommand{\autor}[2]{#1 #2}

\newcommand{\revista}[5]{#1  {\bf#2}, #3  (#5)}
\newcommand{\revistacorto}[4]{#1  {\bf#2}, #3  (#4)}

\newcommand{\editor}[2]{#1 #2}

\newcommand{\y}{{ \rm and\ }}

\newcommand{\libro}[5]{{\it #1} (#3, #4) #5, {#2}}
\newcommand{\librocorto}[4]{{\it#1} (#2, #3) #4}
\newcommand{\chap}{chap.}
\newcommand{\chaps}{chaps.}

\renewcommand{\epsilon}{\varepsilon}
\newcommand{\Tc}{\mbox{$T_C$}}
\newcommand{\DTc}{\mbox{$\Delta\Tc$}}
\newcommand{\Tcx}{\mbox{$\Tc(x)$}}
\newcommand{\kB}{\mbox{$k_{\rm B}$}}
\newcommand{\di}{\mbox{$d_{\rm imp}$}}

\newcommand{\Tmag}{\mbox{$T_{\rm M}$}}

\newcommand{\ie}{{\it i.e.}}
\newcommand{\eg}{{\it e.g.}}

\newcommand{\paper}[1]{Ref.\citeonline{#1}}
\newcommand{\eq}[1]{Eq.(\ref{#1})}
\newcommand{\beq}{\begin{equation}}
\newcommand{\eeq}{\end{equation}}

\newcommand{\gsim}{\stackrel{>}{_\sim}}
\newcommand{\lsim}{\stackrel{<}{_\sim}}

\newcommand{\ns}{\mbox{$n_{\mbox{\it s}}$}}
\newcommand{\nse}{\mbox{$n_{\mbox{\it
s}}(\epsilon)$}}
\newcommand{\nso}{\mbox{$n_{\mbox{\it
s}0}$}}

\newcommand{\Dchi}{\mbox{$\Delta\chi$}}
\newcommand{\Dchie}{\mbox{$\Delta\chi(\varepsilon)$}}
\newcommand{\DchiRKKY}{\mbox{$\Delta\chi^{\rm RKKY}$}}
\newcommand{\DMRKKY}{\mbox{$\Delta\chi^{\rm RKKY}$}}

\newcommand{\chiBT}{\mbox{$\chi_B(T)$}}
\newcommand{\chiB}{\mbox{$\chi_B$}}
\newcommand{\chiT}{\mbox{$\chi(T)$}}
\newcommand{\MBTH}{\mbox{$M_B(T)/H$}}
\newcommand{\MB}{\mbox{$M_B$}}
\newcommand{\MBH}{\mbox{$M_B(H)$}}

\newcommand{\DchiS}{\mbox{$\Delta\chi^{\rm GGL}$}}
\newcommand{\DMS}{\mbox{$\Delta M^{\rm GGL}$}}
\newcommand{\DMSe}{\mbox{$\Delta M^{\rm GGL}(\varepsilon)$}}
\newcommand{\DchiZD}{\mbox{$\Delta\chi^{\rm 0D}$}}
\newcommand{\DMZD}{\mbox{$\Delta M^{\rm 0D}$}}

\newcommand{\DchiMS}{\mbox{$\Delta\chi^{\rm ion\; screening}$}}
\newcommand{\chiBMS}{\mbox{$\chi_B^{\rm magnetic\; ions}$}}
\newcommand{\Dchiind}{\mbox{$\Delta\chi^{\rm mag\,imp}$}}
\newcommand{\Aind}{\mbox{$A$}}

\newcommand{\DM}{\mbox{$\Delta M$}}
\newcommand{\DMe}{\mbox{$\Delta M(\epsilon)$}}
\newcommand{\DMT}{\mbox{$\Delta M(T)$}}
\newcommand{\DMeh}{\mbox{$\Delta M(\epsilon)_{\mbox{\it h}}$}}
\newcommand{\Ds}{\mbox{$\Delta\sigma$}}
\newcommand{\Tsuper}{\mbox{$T^C$}}
\newcommand{\epsilonsuper}{\mbox{$\epsilon^C$}}
\newcommand{\eC}{\epsilonsuper}
\newcommand{\etilde}{\tilde\epsilon}

\newcommand{\Hcdos}{\mbox{$H_{c2}$}}
\newcommand{\Hcdoso}{\mbox{$H_{c2}(0)$}}
\newcommand{\xideodos}{\mbox{$\xi^2(0)$}}
\newcommand{\xideo}{\mbox{$\xi(0)$}}

\newcommand{\citasSyS}{SSuno}
\newcommand{\SySnombre}{Schmidt  and Schmid}

\newcommand{\Matias}{Matthias}

\newcommand{\LOref}{LOuno}
\newcommand{\Kresinref}{Kresin}
\newcommand{\unconventional}{Saxena}


\newcommand{\figurauno}{
\begin{figure}[tb]
\referee{
\mbox{}\hfill\includegraphics[scale=0.53]{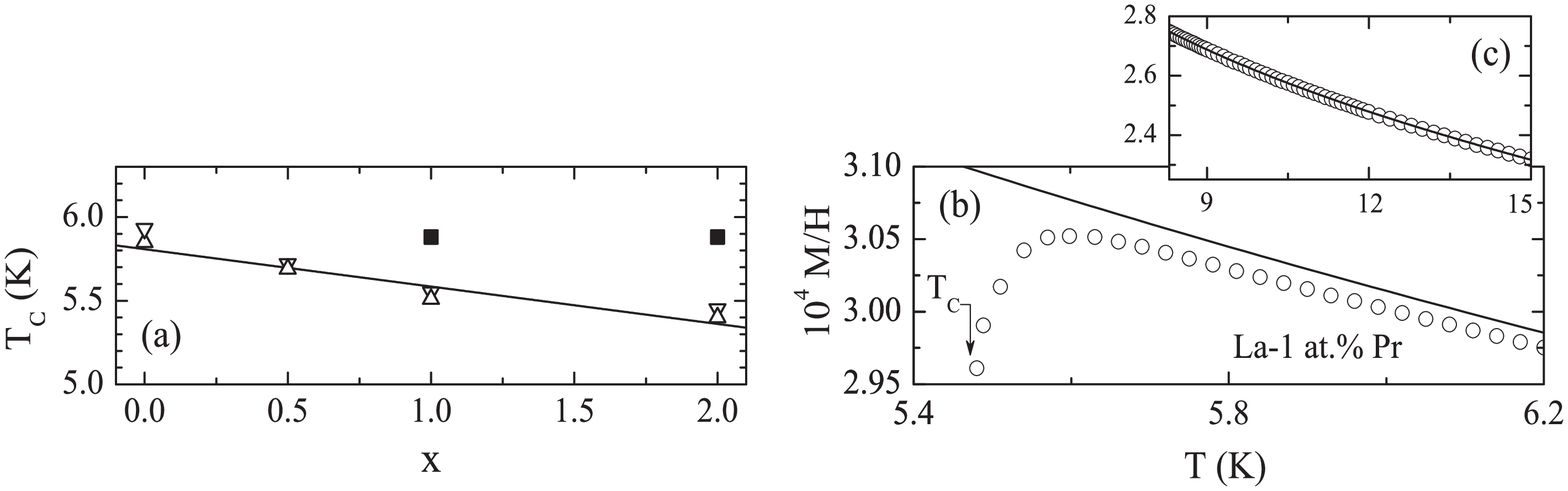}\hfill\mbox{}\\}
\caption{(a) Impurity concentration dependence of the superconducting transition temperature, determined from field-cooled magnetization measurements under $\mu_0H=5\times10^{-4}\mbox{T}$ ($\bigtriangleup$ and \Squaresteel\  data)
 and from the electrical resistivity versus temperature curves ($\bigtriangledown$ data). Solid squares correspond to nonmagnetic (Lu) impurities and open symbols to magnetic (Pr) impurities. The solid line is a fit of the Abrikosov and Gor'kov approach.\cite{AG} (b)~An example, correponding to a La-1$\mbox{}\,\mbox{}$at.\%Pr alloy,  of the as-measured magnetization (over $H$) versus temperature curves, showing directly the differences between the data and the background \MBTH\ (solid line). These data   were obtained under  $\mu_0H=0.05\mbox{T}$, which is well inside the low-field regime for which \DM\ is linear on the field amplitude. The temperature region shown  corresponds well to the reduced temperatures covered in fig.~2(a). (c)~Magnetization over $H$ in the background temperature region for the same alloy.}
\end{figure}
}

\newcommand{\figurados}{
\begin{figure}[tb]
\referee{\mbox{}\hfill\includegraphics[scale=0.5]{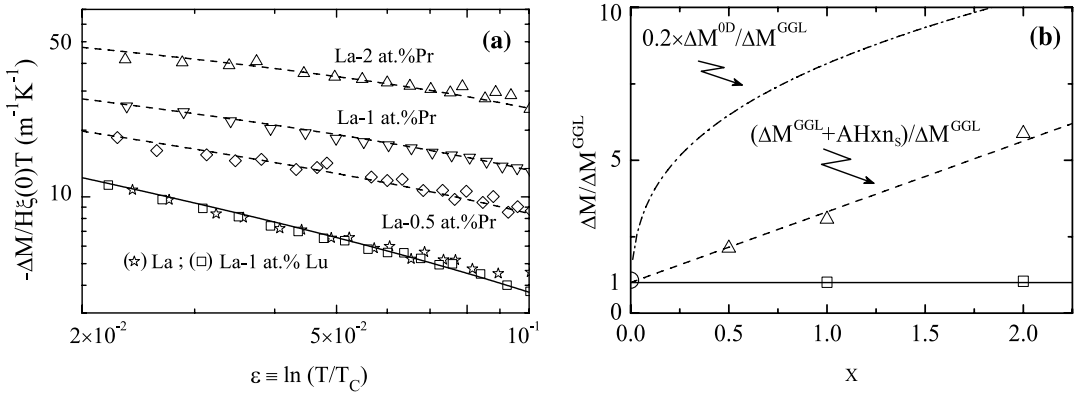}\hfill\mbox{}\\}
\caption{(a) Fluctuation-induced magnetization (over $H\xideo T$) versus reduced temperature in all the La-Pr alloys studied in this work and also for  pure La and a La-1$\mbox{}\,\mbox{}$at.\/\%Lu alloy. These data were obtained under low magnetic fields, $\mu_0H=0.05\mbox{T}$. The covered $\epsilon$-range covered corresponds well to the accessible experimental window: closer to \Tc\ the data are mainly affected by \Tc-inhomogeneities, whereas above $\varepsilon\simeq0.1$ they are mainly affected by background uncertainties. Outside this $\varepsilon$-region the experimental uncertainties could become even bigger than 100\%.  The solid line is the GGL prediction, and the dashed lines the are GGL prediction plus a contribution proportional to the superfluid density and magnetic impurity concentration (see main text for details). (b)~Experimental fluctuation-induced magnetization, relative to the GGL predictions\cite{\citasSyS,Tinkham,MosqueiraPRL}, as a function of impurity content $x$ in pure-La \mbox{(\protect\circulo)}, La-Pr \mbox{($\protect\triangle$)} and La-Lu \mbox{(\protect\cuadrado)}, at $\varepsilon=0.06$ and $\mu_0H=0.05\mbox{T}$. The superconductors without magnetic impurities follow well the GGL  approach (solid line), while in those with magnetic impurities $\left|\DM\right|$ increases, in essence linearly, with $x$. The dashed and dot-dashed lines are explained in the main text.
}
\end{figure}
}

\newcommand{\tablauno}{
\begin{table}
\begin{center}\mbox{}\vskip-0.8cm\mbox{}\\
\begin{tabular}{lccccccc}
\hline
\hline
Sample    &   \Tc  & $\Delta\Tc$ & $\mu_0H_{c2}(0)$ & $\xi(0)$ & $\kappa(\Tc)$&$\ell$&$d_{imp}$ \\
                  &(K)     &(K)                 &(T)                              & (\AA)     &                         &(\AA)&(\AA)\\
\hline
La    &                        5.85    &      0.16      &   0.8       &   200   &   4.4   &   265    & \\
La-0.5 at.\% Pr  &    5.69    &      0.12       &  0.9       &   190    &   5.6   &  150    &    19 \\
La-1 at.\% Pr     &     5.51   &      0.25     &    0.9       &   180   &    6.0   &   110   &    15\\
La-2 at.\% Pr     &     5.40    &     0.18     &    1.0      &    180   &    6.4    &   75     &    12\\
La-2 at.\% Lu     &     5.88   &   0.26     &     0.9     &     190    &  4.2     &  210     &   12\\
\hline
\hline
\end{tabular}
\end{center}
\caption{\label{tabla} Summary of basic parameters of the
La-Pr superconductors studied in this work, obtained from 
magnetization and electrical resistivity measurements.\cite{tobepublished} For comparison, we included also one of
the pure La samples and one of the La-Lu alloys.  The average distance between
impurities $d_{imp}$ was estimated from the nominal sample compositions. $\ell$ is the mean free path of the normal electrons extrapolated to  $T=0 {\rm K}$, and $\kappa(\Tc)$ the Ginzburg-Landau parameter at the critical temperature. The uncertainties in $\xi(0)$, $\kappa(\Tc)$ and $\ell$ are of about  $20\%$.}
\end{table}
}


\newcommand{\titulo}{
Observation of enhanced fluctuation diamagnetism\\ in lanthanum superconductors with\\ dilute magnetic impurities
}

\newcommand{\autores}{F\'elix~Soto, Luc\'{\i}a~Cabo, Jes\'us~Mosqueira,\\ Manuel~V.~Ramallo, Jos\'e~A.~Veira and  F\'elix~Vidal}

\newcommand{\direccion}{
Laboratorio de Baixas Temperaturas e Superconductividade,\footnote{Unidad Asociada al Instituto de Ciencia de Materiales de Madrid, CSIC, Spain}
Departamento de F\'{\i}sica da Materia Condensada,\\
Universidade de Santiago de Compostela, E15782 Spain.
}

\begin{center}
  \Large\bf
\titulo\\  \end{center}\mbox{}\vspace{-1cm}\\

\begin{center}\large\autores\end{center} 

\begin{center}\large\it\direccion\end{center}


\mbox{}\vskip2cm{\bf Abstract. }
The fluctuation-induced diamagnetism \DM, associated with the presence of precursor Cooper pairs in the normal state, has been measured in lanthanum with dilute magnetic (Pr) and nonmagnetic (Lu) impurities. It is found that while for pure La and La-Lu alloys \DM\ agrees, as expected,  with the theoretical predictions, it is much larger for La-Pr alloys (around a factor 5 for La-2$\mbox{}\,\mbox{}$at.\%Pr). These results suggest the existence of an indirect contribution to \DM\  arising from the interaction between fluctuating Cooper pairs and magnetic impurities.

\newpage
\setlength{\baselineskip}{18pt}


The magnetic susceptibility measurements of \Matias\ and coworkers\cite{Matias} showing the decrease of the superconducting transition temperature of lanthanum with dilute magnetic rare earth impurities, and the explanation of these effects by Abrikosov and Gor'kov\cite{AG} in terms of pair-breaking, opened about 45 years ago one of the still at present most interesting and studied issues of  correlated electron systems: the interplay between magnetism and superconductivity.\cite{Fisher,Osborne} A natural question directly related to these pioneering results but to our knowledge not yet addressed until now is how the magnetic impurities will affect the precursor diamagnetism associated with the presence  of fluctuating Cooper pairs created above any superconducting transition by the unavoidable thermal agitation. In fact, this fluctuation-induced diamagnetism above the superconducting transition  may be used as an unique probe to study the competition between pairing correlations and magnetic order without entering into the fully-superconducting state, which would hide the response of the magnetic ions to external fields. 

We present in this Letter measurements of the fluctuation-induced diamagnetism, $\DM\equiv M(T)-M_B(T)$,   in La with dilute (up to 2$\mbox{}\,\mbox{}$at.\%) magnetic (Pr) and nonmagnetic (Lu) impurities.  Here $M(T)$ and $M_B(T)$ are the measured and, respectively, the background or bare magnetizations.
It is found that while for pure La and La-Lu alloys  \DM\ agrees with the calculations on the grounds of the Gaussian-Ginzburg-Landau (GGL) approach  for isotropic 3D superconductors\cite{\citasSyS,Tinkham,Gollub,MosqueiraPRL},  the \DM\ amplitude is much larger in the La-Pr alloys  (about 500\% for impurity concentrations of 2$\mbox{}\,\mbox{}$at.\%). 
The results for La-Lu alloys just confirm previous experiments in other dirty low-\Tc\ superconductors\cite{Tinkham,Gollub,MosqueiraPRL} and they may be easily understood by just assuming that the fluctuating Cooper pairs in the normal state  are also protected, as the Cooper pairs below the superconducting critical temperature \Tc, by the Anderson theorem for symmetric perturbations.\cite{Anderson} These results in La-Lu superconductors provide then a crucial check of the reliability of our measurements of \DM\ in the lanthanum-based alloys. In contrast with these conventional results in La-Lu alloys, the \DM-enhancement in presence of magnetic ions suggests not only that  the fluctuating Cooper pairs in conventional (singlet $s$-wave pairing) BCS superconductors are very robust to antisymmetric perturbations but also the existence of an indirect contribution to \DM\ arising from the interaction between Cooper pairs and magnetic impurities. These results may have implications in other scenarios where superconducting fluctuations and magnetic order coexist\cite{Fisher,Osborne}, in particular when the superconductivity is magnetically mediated\cite{\unconventional}, perhaps including the high-\Tc\ cuprates\cite{Osborne}.  

In addition to its interest as a natural extension of the seminal experiments of \Matias\ and coworkers,\cite{Matias} our choice of the magnetization to study the interplay between fluctuating Cooper pairs in the normal state and magnetic impurities was motivated by the fact that, as known since the early experiments of Tinkham and coworkers in low-\Tc\ superconductors without magnetic impurities,\cite{Tinkham,Gollub} the fluctuation-induced magnetization  probably is the best observable to probe the superconducting fluctuations in any isotropic bulk (3D) superconductor.   In contrast with most of the effects of the superconducting fluctuations on other observables, $\left|\Delta M\right|$ increases  not only with the density of fluctuating Cooper pairs but also with their size, \ie, with the superconducting coherence length amplitude (extrapolated to $T=0\mbox{K}$), \xideo. As a consequence, $\left|\Delta M/M_B\right|$ may take relatively high values at easily accessible temperature-distances from the transition in isotropic low-\Tc\ superconductors, even bigger than in high-\Tc\ superconductors, due to their much larger \xideo.\cite{MosqueiraPRL} In fact, for the superconductors studied in the present Letter we will obtain values of $\left|\Delta M/M_B\right|$ well comparable (about one order of magnitude smaller in the worst case) with those of optimally-doped \mbox{YBa$_2$Cu$_3$O$_{7-\delta}$}.\cite{medidasYBCO} These experimental advantages appear to be crucial when compared with the difficulties that may arise if one uses other observables, very in particular the electrical conductivity, to study the interplay between superconducting fluctuations and magnetic impurities in low-\Tc\ superconductors: mainly in bulk materials, the ratio between the corresponding paraconductivity, \Ds, and the nonfluctuating (background) conductivity, $\sigma_B$, may be orders of magnitude smaller than $\left|\DM/M_B\right|$. In addition,  $\sigma_B$ may be much more affected than $M_B$ by stoichiometric and structural inhomogeneities. These difficulties probably affect deeply the few (to our knowledge) attempts published until now to study through the electrical conductivity the interplay between superconducting fluctuations and magnetic impurities in low-\Tc\ superconductors.\cite{Spahn,Rapp}

  The fitness of the La alloys to measure \DM\ in presence of magnetic impurities is mainly due to the fact that these compounds still have a quite large \xideo, of the order of 200\AA. Complementarily, for some magnetic impurities (very in particular Pr), their normal state magnetization is expected to remain relatively moderate, even under impurity concentrations $x$ as important as 2$\mbox{}\,\mbox{}$at.\%, the larger impurity concentrations used by \Matias\ and coworkers in their  measurements of $\Tc(x)$\cite{Matias}. In addition, at present high-quality La alloys are commercially available. The polycrystalline \mbox{La$_{100-x}$Pr$_x$} and  \mbox{La$_{100-x}$Lu$_x$} alloys studied in this Letter, with $0\leq x\leq 2$, were supplied by Goodfellow and Alfa Aesar, and their impurity concentration was controlled to better than 0.1\%. The superconducting transition width, \DTc, as determined from electrical resistivity and field-cooled magnetic susceptibility measurements remains below $\sim$0.25\/K for all the samples, which confirms their good stoichiometric and structural quality. These measurements, as well as those performed to determine other general characteristics of the samples (very in particular \Hcdoso\ and then \xideo), are detailed elsewhere\cite{tobepublished}. Here we present in  \tabla\ some of the main sample parameters and in fig.~1(a) the impurity concentration dependence of \Tc, which for the La-Pr alloys provides a direct measure of the pair-breaking effects induced by the magnetic impurities. These \Tcx\ data are in agreement with both the results of \Matias\ and coworkers\cite{Matias} and the Abrikosov-Gor'kov approach\cite{AG}.  The solid line in fig.~1(a) is a fit of this approach, which at low concentrations reduces to $\Tc(x=0)-\Tc(x)=\hbar/\kB\tau_\phi(x)$, where $\hbar$ is the reduced Planck constant, \kB\ is the Boltzmann constant and $\tau_\phi(x)=\tau_\phi(x=1)/x$ is the phase pair-breaking time. This leads to $\tau_\phi(x=1)\simeq3.4\times10^{-11}\mbox{s}$, which is much larger than the relaxation time of the normal electrons deduced from resistivity measurements\cite{tobepublished} ($\tau\sim10^{-14}\mbox{s}$). \Tcx\ for the alloys with nonmagnetic Lu-impurities is almost constant, the Cooper pairs in these last alloys being protected, as is well known, by the Anderson theorem.\cite{Anderson}

An example of the temperature behaviour above \Tc\ of  the measured and of the  background magnetizations  is presented in figs.~1(b) and (c).  These measurements were performed under constant magnetic field amplitude (and in the low-field regime,  $H\ll\Hcdoso$) with a commercial, SQUID based, magnetometer  (Quantum Design's MPMS) and by using cylindrical samples of diameter  and height $\sim5\mbox{mm}$,  the maximum volume allowed by our magnetometer. The demagnetizing effects are negligible above \Tc. Other experimental details are similar to those  in magnetization measurements in other low-\Tc\ superconductors.\cite{MosqueiraPRL} 
The solid line in figs.~1(b) and (c) is the background or bare magnetization (over $H$), obtained  by fitting in the temperature region bounded by $1.5\Tc\leq T\leq3\Tc$ the function $\MBTH=\chi_0+BT+C/T$ (with $\chi_0$, $B$ and $C$ as free parameters). In this region the agreement between \MBTH\ (solid line) and the data is excellent, the maximum deviation being below 0.05\%. Let us stress that even above this background fitting region, up to $5\Tc$, the agreement with the data still remains quite good, the maximum deviation being below 5\%.
 The above background functionality is expected to be a good approximation  for (anti)ferromagnetic materials at  temperatures well above the Curie or N\'eel temperatures \Tmag,\cite{libromagnetismo} which in superconductors with diluted magnetic impurities may be estimated through the relationship\cite{deGennes} $\Tmag(x)\sim\Tc(x=0)-\Tc(x)$. In our samples this leads to $\Tmag\lsim0.5\mbox{K}\ll\Tc$.

From curves similar to those presented in fig.~1(b), we obtain   the fluctuation-induced magnetization versus  reduced-temperature  curves, \DMe, where $\varepsilon\equiv\ln(T/\Tc)$. In fig.~2(a) we present some examples of these \DMe\ curves, very in particular those that correspond to the La-Pr alloys. These data are well inside the $\varepsilon$-range where the GGL approach is expected to be applicable.\cite{\citasSyS,Tinkham,Gollub,MosqueiraPRL,medidasYBCO} The solid line corresponds to the GGL calculations for  isotropic 3D superconductors without magnetic impurities  in the zero-field limit,\cite{\citasSyS,Tinkham,MosqueiraPRL}  
\begin{equation}
\frac{\DMSe}{H\xideo T} = 
 \frac{-\mu_0\kB}{3\phi_0^2} 
 \left[
 \frac{{\rm arctan}\sqrt{\frac{\mbox{$\varepsilon^C-\varepsilon$}}{\mbox{$\varepsilon$}}}}
 {\sqrt{\varepsilon}}
 -
 \frac{{\rm arctan}\sqrt{\frac{\mbox{$\varepsilon^C-\varepsilon$}}{\mbox{$\varepsilon^C$}}}}
 {\sqrt{\varepsilon^C}}
 \right],
 \label{DMSS}
\end{equation}
where $\eC\simeq0.6$ is the BCS value of the cutoff constant\cite{EPLVidal}  and $\phi_0$ is the flux quantum.  As visible in this figure, the data in pure La and in La-Lu alloys agree with \eq{DMSS} well within the experimental uncertainties, which are of the order or below 40\%\cite{cuarenta}, an agreement that was already observed in other low-\Tc\ superconductors with nonmagnetic impurities.\cite{MosqueiraPRL} As stressed before, this result may be easily understood by just assuming the applicability also above \Tc\ of the immunity predicted by Anderson\cite{Anderson} of the Cooper pairs to symmetric perturbations. So,   these last results also show the reliability of our present experiments to probe the superconducting fluctuations above \Tc\ in La alloys.

The results of fig.~2(a) show also that in presence of magnetic impurities {\it both} the {\it amplitude}  and the {\it $\varepsilon$-dependence} of $\DMe/H\xideo T$  differ from the GGL calculations\cite{\citasSyS,Tinkham,MosqueiraPRL}, even for the alloy with the lower Pr concentration: The $\DMe/H\xideo T$ amplitudes remain much larger, about five times for the La-2~at.\%Pr alloy, and also their reduced-temperature dependence is smoother. As illustrated quantitatively  in fig.~2(b), this enhancement is essentially proportional to the magnetic impurity concentration, in striking contrast with the well established and well understood decrease of the superconducting transition temperature when such a concentration increases\cite{Matias,AG,Fisher} (see fig.~1(a)).

 In analyzing the experimental findings summarized in figs.~2(a) and (b) we must start by checking if such an strong \DM-enhancement is just an effect due to \Tc-inhomogeneities at long-length scales (bigger than \xideo), in turn associated with  stoichiometric  and structural defects.\cite{Vidalimpurezas} However, this extrinsic effect is ruled out not only by the high quality of the samples but mainly by the fact that the La-Lu alloys display magnetic and electrical resistivity transition widths, \DTc, similar to those of their La-Pr counterparts (see \tabla). Also, a rounding of the magnetic transition due to local \Tc-inhomogeneities, induced by the magnetic ions (inhomogeneities either isolated\cite{\LOref} or Josephson tunneling-connected\cite{\Kresinref}), may be discarded because in our La-Pr alloys the mean distances among impurities, \di, are in the range $12-19$\AA, which is well below the \xideo\ values $\sim200$\AA\ (see \tabla).

The amplification effects associated with a  reduction of the superconducting fluctuations' dimensionality (in turn due to a confinement of these fluctuations  between the impurity interspaces) may be also discarded. This is illustrated  in fig.~2(b), where the dot-dashed line corresponds to the fluctuation magnetization that may be obtained by supposing that the magnetic impurities would allow only 0D superconducting fluctuations occurring in the interspace between magnetic impurities. For such a \DMe\ in 0D we used\cite{Tinkham}
$
\DMZD(\epsilon)=(\xideo/\di\varepsilon^{1/2})\DMSe.
$

Another possible amplification effect, opposite in origin to a superconducting fluctuations' confinement, would be that these fluctuations embrace part of the magnetic ions, cancelling their contribution to the magnetization. A crude estimation of these effects may be done by just identifying $\left|\DMS/H\right|$ with the effective volume fraction occupied by the evanescent superconducting ``droplets''. The resulting contribution to the magnetization is then $\MB\DMS/H$ which, taking into account that  in our samples $10^{-4}\lsim\MB/H\lsim10^{-3}$ near the transition, leads to a negligible effect.  Finally, we also note that a mechanism of internal-field screening different to Meissner-like currents, which involves spin alignment of the quasiparticles composing the Cooper pair, has been recently proposed, for $T<\Tc$ and noninteracting magnetic impurities.\cite{EPL} However, when applied to $T>\Tc$ the resulting change in magnetization  is again about three orders  of magnitude smaller than our observations.

In absence of a theoretical approach for the superconducting fluctuations in presence of magnetic impurities, and as a {\it first crude} attempt to establish {\it empirically} the physical origin of the observed \DM-enhancement, we have checked if it is possible to  separate the measured \DMe\ in two contributions: the direct GGL \SySnombre\ term and an indirect contribution proportional to both the density of fluctuating Cooper pairs and the magnetic impurity concentration. This check has been summarized in figs.~2(a) and 2(b), where the dashed lines were obtained  using the empirical expression 
 \begin{equation}
 \DMe=\DMSe+\Aind \, H \, x \, \nse, 
 \label{empirico}
 \end{equation}
where \Aind\ is a constant and \nse\ is the superfluid density  (in dimensionless units)  for isotropic 3D superconductors  in the zero-field limit,\cite{EPLVidal} 
 \begin{equation}
\nse
\hskip-0.2em=\hskip-0.2em
\frac{\kB \mu_0 {\rm e}^2 T\xideo}{\hbar^{2}\sqrt{\varepsilon}} \left[
\sqrt{\frac{\varepsilon^C-\varepsilon}{\varepsilon}}
-
{\rm arctan}\sqrt{\frac{\varepsilon^C-\varepsilon}{\varepsilon}}
\right]\hskip-0.2em.
\label{eqns}
 \end{equation}
   The dashed lines in fig.~2(a) correspond to fits to the La-Pr alloy data of the above $\DMe$ expression, with \Aind\ as the only free parameter (we constrained the rest of parameters to be compatible with \tabla, and also used $\varepsilon^C=0.6$ for all the fits). As it may be seen in this figure,  the agreement between the experimental results and \eq{empirico} is excellent, for {\it both} the amplitude {\it and} the $\varepsilon$-behaviour, and this using the same  value $\Aind\simeq-0.75$ for all the samples.
   
   The above results lead to two complementary conclusions: First, the presence of the direct  GGL \SySnombre\ contribution in the La-Pr alloys suggests the robustness of the fluctuating Cooper pairs to dilute magnetic impurities. Although unexpected when compared to the decrease of \Tcx\ shown in fig.~1(a), this behaviour agrees with recent \DM-measurements in Pb-In alloys under high reduced-fields,\cite{Soto} which also suggests the robustness of the fluctuating Cooper pairs to the antisymmetric nature of an applied magnetic field.   Complementarily, the presence of an indirect contribution to  \DMe\  in the La-Pr alloys suggests that the fluctuating Cooper pairs modify the coupling between the magnetic ions. In the case of dilute alloys such a coupling is mediated by the electronic sea (RKKY model).\cite{libromagnetismo,RKKY} It may be then expected that the changes induced in the electrons' spin susceptibility by the Cooper pair formation will also affect the magnetization due to the magnetic ions. In fact, such an effect was early proposed below \Tc\  \cite{GRFF}, and the corresponding change in the magnetization above \Tc\ may be expected to be proportional to both the superfluid density and the concentration of magnetic ions, in agreement with our experimental findings. These results suggest that the energetic balance between the magnetic and the superconducting orders could be shifted by the presence of precursor Cooper pairs. Although these crude ideas need to be confirmed both experimentally and theoretically, the experimental findings presented in this Letter already stress that the superconducting fluctuations may play an unexpected important role in the interplay between magnetism and superconductivity, mainly if  the latter  is magnetically mediated.\cite{Osborne,Saxena}

\mbox{}

{\bf Acknowledgments}

We acknowledge useful conversations and correspondence with J.B.~Goodenough and K.~Maki on this topic. This work was supported by the CICYT, Spain (grant no.\ MAT2004-04364), the Xunta de Galicia (PGIDT01PXI20609PR), and Uni\'on Fenosa (220/ 0085-2002). LC acknowledges financial support from Spain's Ministerio de Educaci\'on y Ciencia through a FPU grant.

\mbox{}

\referee{
\newpage\mbox{}\\ \figurauno \mbox{}\\
\newpage\mbox{}\\ \figurados \mbox{}\\
\newpage\mbox{}\\ \tablauno\mbox{}\\ }


\begin{thebibliography}{99}
\bibitem{cuarenta}These uncertainties are mainly due to the background extraction. They have been estimated by varying by 10\% the temperature location of the background fitting region.
\end{thebibliography}
\end{document}